\documentstyle[mprocl]{article}

\catcode`\@=11
\@addtoreset{equation}{section}
\catcode`\@=12
\begin{document}
\baselineskip8mm
\title{\vspace{-6cm}Complex inflaton field in quantum cosmology}
\author{A. Yu. Kamenshchik$^{1 \dag}$, \ I. M. Khalatnikov$^{1,2
\dagger}$ and \ A. V. Toporensky$^{3 *}$}
\date{}
\maketitle
\hspace{-6mm}$^{1}${\em L.D. Landau Institute for
Theoretical Physics, Russian
Academy of Sciences, Kosygin str. 2, Moscow, 117334, Russia}\\
$^{2}${\em Tel Aviv University,
Tel Aviv University, Raymond and Sackler
Faculty of Exact Sciences, School of Physics and Astronomy,
 Ramat Aviv, 69978, Israel}\\
$^{3}${\em Sternberg
Astronomical Institute, Moscow University, Moscow, 119899, Russia}\\
\\

We investigate the cosmological model with the complex scalar
self-interacting inflaton field non-minimally coupled to gravity.
The different geometries of the Euclidean classically forbidden
regions are represented. The instanton solutions of the corresponding
Euclidean equations of motion are found by numerical calculations
supplemented by the qualitative analysis of Lorentzian and Euclidean
trajectories.
The applications of these solutions to the no-boundary and tunneling
proposals for the wave function of the Universe are studied.
Possible interpretation of obtained results and  their
connection with inflationary cosmology is discussed. The restrictions
on the possible values of the new quasi-fundamental constant of the
theory--non-zero classical charge-- are obtained.
The equations
of motion for the generalized cosmological model with complex scalar
field are written down and investigated. The conditions of the
existence of instanton solutions corresponding to permanent values
of an absolute value of scalar field are obtained.
\\
\\
$^{\dag}$ Electronic mail: kamen@landau.ac.ru\\
$^{\dagger}$ Electronic mail: khalat@itp.ac.ru\\
$^{*}$ Electronic mail: lesha@sai.msu.su\\

\section{Introduction}
\hspace{\parindent}
It is widely recognized that
inflationary cosmological models give a good basis for the
description of the observed structure of the Universe $^{1}$.
Most of these models include the so called inflaton scalar field
possessing non-zero classical average value which provides the
existence of an effective cosmological constant on early stage of
the cosmological evolution. On one hand inflationary cosmology
has received the strong support due to discovery of the anisotropy
of the microwave background radiation $^{2}$ while on the other hand
it is connected with such an exiting field of modern theoretical
physics as quantum cosmology. The main task of quantum cosmology
is the consideration of the Universe as a unique quantum object
which can be described by the wave function of the Universe,
obeying to the Wheeler-DeWitt equation. Studying this wave function
of the Universe one can hope to get the probability distribution
of the initial conditions for the Universe.

During the last decade quantum cosmology has been developing
intensively on the basis of two proposals for the boundary conditions
for the wave function of the Universe: the so-called ``no-boundary''
$^{3}$ and ``tunneling'' $^{4}$ proposals. Both  these proposals use the
apparat of Euclidean quantum field theory combined with the ideas of
the theory of quantum tunneling transitions and instantons. However,
these proposals taken in tree semiclassical approximation cannot
provide the normalizability of the wave function of the Universe $^{5}$
and to predict the initial conditions for the cosmological evolution
providing sufficient amount of inflation $^{6}$.

One can look for different ways out from this situation.
Consideration of the wave function of the Universe in one-loop
approximation $^{7}$ gives us an opportunity to obtain the
normalizability of the wave function of the Universe and the
existence of the suitable probability distribution for the initial
conditions for inflation provided proper particle content of the
theory is chosen.

Another possible direction of the development of quantum cosmology
is the consideration of more wide theories than traditional scheme
with a real scalar field. Thus in the series of recent papers $^{8,9}$
the model with a complex  scalar inflaton field was studied.
One of the reasons for the consideration of a complex scalar field
consists in the fact that such fields and the non-Abelian multiplets
of scalar fields appear naturally in the modern theories of particle
physics.

The most natural representation of the complex scalar field has the
form
\begin{equation}
\phi = x \exp (i \theta),
\end{equation}
where $x$ is the absolute value of the complex scalar field while
$\theta$ is its phase. This phase is cyclical variable corresponding
to the conserved quantity -- a classical charge of the Universe,
which plays the role
of the new  quasi-fundamental constant of the theory $^{8}$.

Appearance of this new constant essentially modifies the structure of
the Wheeler-DeWitt equation. Namely, the form of the superpotential
$U(x,a)$, where $a$ is the cosmological radius of the Universe
displays now a new and interesting feature:  the Euclidean region,
i.e. the classically forbidden region where $U > 0$, is bounded by a
closed curve in the minisuperspace $(x, a)$ for a large range of
parameters.  Thus, in contrast with the picture of ``tunneling from
nothing''$^{4}$ and with the ``no-boundary proposal'' for the wave
function of the Universe $^{3}$ we have Lorentzian region at the very
small values of cosmological radius $a$ and hence a wave can go into
Euclidean region from one side and outgo from the other.  These new
features of the model require reconsideration of traditional schemes
$^{3,4}$ and give some additional possibilities.

Here, it is necessary to stress that speaking about the ``Euclidean''
or ``classically forbidden'' regions one should understand that in
the case of the quantum gravity and cosmology these terms can be used
only in ``loose'' sense, because due to indefiniteness of the
supermetric the ``Euclidean'' region is not impenetrable for
Lorentzian trajectories. It is well-known that in the cosmological
models with inflaton scalar field the Lorentzian trajectories can
penetrate into Euclidean region just like as the trajectories
corresponding to the Euclidean equations of motion can leave
Euclidean region for the Lorentzian one (see, for example, Refs.
$^{10 -- 12}$. However, one can use such terms as Euclidean and
Lorentzian regions as people usually do investigating the tunneling
type processes in cosmology $^{3--12}$ and in instanton physics
$^{13}$ even in the case of indefinite supermetric. Moreover, one can
give to the term ``Euclidean region'' quite definite value as the
region where the points of the minimal contraction and maximal
expansion of the Universe can exist (cf. Sec. V of the present
paper).

In principle, investigating the cosmological models with the complex
scalar field one can use instead of parametrization (1.1) the pair of
real scalar fields, representing real and imaginary part of $\phi$
as it was done in $^{14}$. However, in this case the presence of the
symmetry corresponding to new constant is hidden and results obtained
in $^{14}$ for the case of flat Freedman-Robertson-Walker model
coincide with those obtained for the case of the real scalar field
 $^{11}$.

In recent years quite a few papers were devoted to the
investigation of the cosmological models with the non-minimal
coupling between inflaton scalar field and gravity $^{15,7}$. On one
hand such models give a lot of opportunities for matching with the
observational data and ideas of particle physics, on the other hand
they can be treated as more consistent from the point of view of
quantum gravity $^{16}$.

In our recent letter $^{17}$, we have considered non-minimally coupled
complex scalar field. It was shown that inclusion of non-minimal
coupling makes the model more rich. In particular the geometry of
Euclidean regions have a large diversity and depends on the choice of
parameters of the theory.  In the model with the minimally coupled
complex scalar field it was shown that one instanton solution exist
$^{8}$.  This instanton solution can be continued into the Lorentzian
region in accordance with Lorentzian equations of motion and thus to
provide the beginning of the inflation. This scheme gives the strong
preference to the ``no-boundary'' proposal for the wave  function of
the Universe$^{9}$. At the same time in the case of non-minimally
coupled complex scalar field we can have a couple of instantons one
of which is suitable for the ``no-boundary'' wave function of the
Universe while the other is suitable for the tunneling one. This
paper will be devoted to the more detailed investigation of the
properties of the model represented in $^{17}$.

Here it is necessary to tell that strictly speaking the
Hartle-Hawking or ``no-boundary''proposal can not be literally applied
to the case of complex scalar field$^{9}$. Indeed, in the
semiclassical approximation and in a minisuperspace model
``no-boundary''proposal can be boiled down to the path integration
from $a=0$ up to some small value of the cosmological radius $a$ over
compact metrics and regular matter fields. However, in the presence
of the phase variable $\theta$ and connected with it new constant--
charge of the Universe -- the centrifugal term arises in the
super-potential which prevents regularity of matter fields at $a=0$.
Thus we need in some extention of the Hartle-Hawking proposal.
Throughout this paper we shall use the extention of no-boundary
proposal discussed in $^{9}$. The idea consists in the consideration
that Hartle-Hawking proposalis equivalent to the requirement that the
wave function of the Universe should be exponentially growing in the
classically forbidden region and should be in semiclassical
approximation proportional to
\[\exp(-I), \]
where $I$ is the corresponding classical Euclidean action. Instantons
considered in Refs. 8,9,17 corresponds to just this definition of the
``no-boundary'' wave function of the Universe.

It is necessary to add that another approach to the complex scalar
field in cosmology was discussed in $^{18--21}$ mainly in the context
of wormhole solutions. It was noticed $^{19}$, that ``nonzero charge
can play the same role in wormhole dynamics that nonzero angular
momentum does in the dynamics of a particle in an attractive central
potential, or nonzero magnetic charge in the dynamics of an 't Hooft
-Polyakov monopole. Nonzero angular momentum keeps the particle from
falling into the origin; nonzero magnetic charge keeps the monopole
from decaying into mesons; nonzero charge flowing down the throat
keeps the wormhole from punching off two disconnected manifolds''.
Some differences between our approach and that of Refs. $^{18--21}$
will be discussed below.

The structure of our paper is as follows: in Sec. II we obtain
the new fundamental constant of our model -- classical charge --and
discuss its influence on the structure of the superpotential and
on the tunneling process; in Sec. III we consider the different forms
of Euclidean region depending on the choice of parameters; in Sec. IV
we present the equations of motion for our theory and describe the
results of numerical search of instantonic solutions of Euclidean
equations of motion and their connection with different versions
of the boundary conditions for the wave function of the Universe;
in Sec. V we present the equations of motion for the generalized
model of the complex scalar field interacting with gravity and study
their properties; Sec. VI is devoted to the brief summary of the
obtained results.

\section{Complex scalar field and the new quasi-fundamental constant}
\hspace{\parindent}
We shall consider the model with the following action:
\begin{eqnarray}
&& S = \int d^{4}x \sqrt{-g}\left(\frac{m_{P}^{2}}{16 \pi} (R - 2
\Lambda) + \frac{1}{2} g^{\mu \nu} \phi_{\mu}^{*} \phi_{\nu} \right.
\nonumber  \\
&&\left. + \frac{1}{2} \xi R \phi \phi^{*}- \frac{1}{2} m^{2} \phi
\phi^{*} - \frac{1}{4!} \lambda (\phi \phi^{*})^{2}\right).
\end{eqnarray}
Here $\xi$ is the parameter of non-minimal coupling  (we choose for
convenience the sign which is opposite to the generally accepted),
$\lambda$ is the parameter of the self-interaction of the scalar
field, $\Lambda$ is cosmological constant, $m$ is the mass of the
scalar field. The complex scalar field $\phi$ can be represented in
the form given by Eq. (1.1).

We shall consider the minisuperspace model with the spatially
homogeneous variables $a$ (cosmological radius in the Freedman-
Robertson-Walker metric), $x$ and $\theta$.
In terms of these variables the action (2.1) looks as follows:
\begin{eqnarray}
&& S = 2\pi^{2}\int dt N a^{3}\left(\frac{m_{P}^{2}}{16 \pi}
\left[6\left(\frac{\dot{a}^{2}}{N^{2}a^{2}}+\frac{\ddot{a}}{N^{2}a}+
\frac{1}{a^{2}}\right) - 2
\Lambda\right] + \frac{1}{2N^{2}} \dot{x}^{2} \right.  \nonumber  \\
&&\left. + \frac{1}{2N^{2}} x^{2} \dot{\theta}^{2}+ 3 \xi
\left(\frac{\dot{a}^{2}}{N^{2}a^{2}}+\frac{\ddot{a}}{N^{2}a}+
\frac{1}{a^{2}}\right)x^{2}-
\frac{1}{2} m^{2} x^{2}
- \frac{1}{4!} \lambda x^{4}\right),
\end{eqnarray}
where $N$ is the lapse function.
Now, by integrating by parts one may get rid of the terms containing
$\ddot{a}$ and write down the action (2.2) in more convenient form
\begin{eqnarray}
&& S = 2 \pi^{2}\int dt N \left(\frac{m_{P}^{2}}{16 \pi}
\left[6\left(-\frac{\dot{a}^{2}a}{N^{2}}+
a\right) - 2
\Lambda a^{3}\right] + \frac{1}{2N^{2}} \dot{x}^{2} a^{3}
+\frac{1}{2N^{2}} x^{2} \dot{\theta}^{2} a^{3}
\right.  \nonumber  \\ &&\left.
+ 3 \xi \left(\frac{-\dot{a}^{2} a}{N^{2}}
+a\right)x^{2}-6 \xi \frac{\dot{a} \dot{x} a^{2} x}{N^{2}}
-\frac{1}{2} m^{2} x^{2} a^{3}
- \frac{1}{4!} \lambda x^{4} a^{3} \right).
\end{eqnarray}

Let us notice that the phase variable $\theta$ is the cyclical one
and correspondingly, its conjugate momentum $p_{\theta}$ should be
conserved. We shall call its value by a charge of
the Universe and  shall denote it by $Q$
\begin{equation}
p_{\theta} = Q = a^{3} x^{2} \dot{\theta}.
\end{equation}

Now, coming to the canonical formalism and using the relation
(2.4) one can rewrite the action (2.3) in the following form:
\begin{equation}
S = 2\pi^{2} \int dt (p_{a} \dot{a} + p_{x} \dot{x} - N {\cal H}),
\end{equation}
where super-Hamiltonian ${\cal H}$ has the following form
\begin{eqnarray}
&&{\cal H} =  -\frac{p_{a}^{2}}{24 a \left(\frac{m_{P}^{2}}{16\pi}
+\frac{\xi x^{2}}{2} + 3 \xi^{2} x^{2}\right)}
-\frac{\xi p_{x} p_{a} x}{2 a^{2} \left(\frac{m_{P}^{2}}{16\pi}
+\frac{\xi x^{2}}{2} + 3 \xi^{2} x^{2}\right)}\nonumber \\
&&+\frac{p_{x}^{2}}{2 a^{3}} \frac{\left(\frac{m_{P}^{2}}{16\pi}
+\frac{\xi x^{2}}{2}  \right)}
{\left(\frac{m_{P}^{2}}{16\pi}
+\frac{\xi x^{2}}{2} + 3 \xi^{2} x^{2}\right)}
-U(a,x).
\end{eqnarray}
The function $U(a,x)$ which we shall call the superpotential
looks as follows:
\begin{eqnarray}
&&U(a,x) = a
\left(\frac{m_{P}^{2}}{16 \pi}(6 - 2 \Lambda a^{2})) + 3 \xi
x^{2}\right. \nonumber \\ &&\left.  -\frac{Q^{2}}{a^{4} x^{2}} -
\frac{1}{2} m^{2} x^{2} a^{2} - \frac{1}{24} \lambda x^{4}
a^{2}\right).
\end{eqnarray}
The variation of the action in
respect to lapse function $N$ gives us the super-Hamiltonian
constraint
\begin{equation}
{\cal H} = 0,
\end{equation}
whose quantum analog is the well-known Wheeler-DeWitt equation.

Now it is convenient to go back to Lagrangian formalism and to write
down the Lagrangian, depending only on two minisuperspace variables
$a$ and $x$ and their derivatives (the lapse function $N$ is chosen
to be equal to 1):
\begin{eqnarray}
&& L = \left(\frac{m_{P}^{2}}{16 \pi}
\left[6\left(-\dot{a}^{2}a+
a\right) - 2
\Lambda a^{3}\right]
\right.  \nonumber  \\ &&\left.
+ 3 \xi \left(-\dot{a}^{2} a
+a\right)x^{2}-6 \xi \dot{a} \dot{x} a^{2} x
+\frac{1}{2} \dot{x}^{2} a^{3}\right.\nonumber \\
&&\left.-\frac{Q^{2}}{2 a^{3} x^{2}}
-\frac{1}{2} m^{2} x^{2} a^{3}
- \frac{1}{4!} \lambda x^{4} a^{3} \right).
\end{eqnarray}
This Lagrangian gives us the equations of motion for the
minisuperspace variables $a$ and $x$ which will be written down and
investigated in Sec. IV. (Let us notice that the simple
substitution of the value of $\dot{\theta}$ from Eq. (2.4) with the
subsequent variation of (2.2) in respect to $a$ and $x$ does not
gives us the correct equations of motion).

It is clear that the transition from the Lorentzian equations of
motion which can be obtained from the Lagrangian (2.9) to their
Euclidean counterparts should be done by simple change of sign before
terms containing time derivatives of $a$ and $x$. However, if we make
the transition to Euclidean (imaginary) time {\it before} we have
got rid of $\dot{\theta}$ due to Eq. (2.4) we shall have {\it
another} couple of Euclidean equations of motion. These equations
of motion shall differ from the previous one by the sign before
terms containing the new charge $Q$. The question arises: what pair
of equations is ``correct'' and what is the reason of the
``discrepancy''? These problems were discussed in $^{18}$ and
illustrated by the example of the particle in the central potential,
where the angular momentum plays the role of charge while cyclical
angular variable plays the role of phase. Assuming that the correct
Euclidean equations of motion are those which are obtained when the
transition to Euclidean time is carried out {\it after} the
exclusion of cyclic variables author of Ref. 18 proposed two ways
to cure the situation.

One proposal consists in the idea that it is necessary to add to the
Euclidean action where all the variables are on equal footing the
term which representing Lagrange multiplier multiplied by
constraint
\begin{equation}
\frac{d Q}{d \tau},
\end{equation}
where $\tau$ is Euclidean time. Indeed, the addition of the term
proportional to (2.10) to the action changes the Euclidean equations
of motion in a desirable way. However, to our mind
the very procedure of adding of (2.10) is an unnecessary and
illegitimate simultaneously. It is unnecessary, because the
Lagrangian of the theory implies the conservation of charge (i.e.
the momentum conjugate to cyclic variable). Moreover, it is
illegitimate because, the term (2.10) contains the second derivative
of our variable $\theta$ and its addition to the action is equivalent
to the forceful change of sign before terms containing $Q$.
(One can say also that if the procedure of addition of (2.10) is
legitimate then analogous procedure can be carried out with
Lorentzian equations of motion as well and we shall have apparently
incorrect signs before some terms in these equations).

Another idea uses the known rules for matching the solutions
of classical Lorentzian equations of motion with their Euclidean
counterparts $^{12,22}$. To satisfy the principle of minimal action
it
is necessary not only to require the dynamical variables
obey to Lorentzian and Euclidean equations of motion along
the Lorentzian and Euclidean sections of their trajectories
 correspondingly, but also to choose the point of matching of
Lorentzian and Euclidean trajectories in such a way to provide
the vanishing of the first derivatives of these variables in this
point. If it is impossible, then the complexification of the
trajectories is inescapable and in the points of matching the
following conditions should be satisfied:
\begin{equation}
Re \dot{q}_{E} = Im \dot{q}_{L}; \;\;
Im \dot{q}_{E} = - Re \dot{q}_{L}.
\end{equation}
Now looking at the Eq. (2.4) it is easy to understand that
$\dot{\theta}$ cannot vanish and thus if we want to include
$\theta$ into the set of our minisuperspace variables and to treat it
on equal footing with $x$ and $a$ then the complexification at least
of the phase variable is necessary. In this case at coming to
Euclidean region and at the transition from Lorentzian equations of
motion to Euclidean ones it is necessary using the conditions (2.11)
to go from the real phase $\theta$ to the complex one having
imaginary component. Carrying this procedure out simultaneously with
the transition to Euclidean (imaginary) time we again obtain the
Euclidean equations of motion which coincide with those obtained
from the effective Lagrangian (2.9) including only two variables
$a$ and $x$.

Summing up one can say that Euclidean equations of motion which
are obtained from the Lorentzian one {\it before} exclusion
of phase variable $\theta$ coincide with those obtained {\it after}
exclusion $\theta$ provided the complexification of phase was carried
out. Such procedure is quite correct from the mathematical point of
view, however the appearance of an imaginary phase can evoke some
difficulties connected with the interpretation.

>From our point of view there is not necessity to consider the
imaginary phases at the transition to the Euclidean equations of
motion. It is much more reasonable to think that the phase variable
$\theta$ does not subject to the transition from the Lorentzian time
to the Euclidean time and during the tunneling transition continues
to ``live'' in Lorentzian time. In the framework of such an approach
we naturally obtain the effective Lagrangian (2.9) and the
corresponding Euclidean equations of motion. This approach accepting
the simultaneous evolution of different variables in Lorentzian and
Euclidean times can find justification in the fact that the very idea of
introduction of Euclidean equations of motion and instantons is
connected with the impossibility to describe the process of tunneling
in terms of classical trajectories. However, not all the variables of
the quantum system which undergoes the process of tunneling should be
treated on equal footing. Some of them are not involved into the
process of tunneling and their evolution can be described in terms of
Lorentzian equation of motion. This fact is well-known in the
non-relativistic quantum mechanics $^{23}$. Moreover, the existence
of the variables which do not undergo the tunneling transition while
the system in whole does, can be used for the definition and
measurement of time spent by system under barrier $^{24}$, because
these ``non-tunneling'' variables can be used as a ``quantum clock''
$^{25}$.

The problem consists in the fact that there is not a procedure of
subdividing of the degrees of freedom of the system under
consideration onto ones which undergo tunneling and those which do
not. This problem is rather complicated even in non-relativistic
multidimensional quantum mechanics let alone the quantum gravity,
where one additionally encounters the problem of indefiniteness of
supermetric. However, in the situation when we have cyclical
variables the problem becomes more simple because, we can express
these variables through other ones and conserved quantities reducing
in such a way the problem of tunneling to that of lower
dimensionality.

Thus, in the rest of our paper we shall think that the phase $\theta$
does not undergo through tunneling and can be safely expressed
through new quasi-fundamental constant - classical charge $Q$ and
variables $x$ and $a$ through relation (2.4). Dynamics of these
variables is described by the Lagrangian (2.9) and will be studied in
Sec. IV. In the next section we shall consider the geometry of the
so-called Euclidean or ``classically-forbidden'' regions
which can be determined as the regions of the positivity of the
superpotential $U(a,x)$.

\section{Geometry of Euclidean regions}
\hspace{\parindent}
We have already mentioned that the very notion of the Euclidean
region for the multidimensional problems and especially in the
problems with the indefinite supermetric as in the case of quantum
gravity and cosmology becomes ``fuzzy''. Nevertheless, we shall use
this terminology, because it is generally accepted. Moreover, we
shall show in the Sec. V that the notion of Euclidean region has the
quite well-defined physical sense.

Thus we shall call Euclidean region that one where the following
condition takes place:
\begin{equation}
U(a,x) > 0.
\end{equation}
Correspondingly the boundary of this region is given by the equation
\begin{equation}
U(a,x) = 0.
\end{equation}

Resolving this equation we can get the form of the Euclidean
region in  the plane of minisuperspace variables $(a,x)$. It is
interesting to compare the form of these regions for different values
of parameters included on the superpotential $U$ (Eq. (2.7)).

In the simplest case then $Q = \Lambda = \lambda = \xi = 0$ we have
non-compact Euclidean region bounded by hyperbolic curve $x = \pm\sqrt
{\frac{3}{4 \pi}} \frac{m_{P}}{m a}$ (see Fig. 1a).
Inclusion of the cosmological term $\Lambda \neq 0$ implies the
closing of the Euclidean region ``on the right'' at $a =
\sqrt{\frac{3}{\Lambda}}$ (see Fig. 1b).

Inclusion of the non-zero
classical charge of the scalar field $Q \neq 0$ implies the closing
of the Euclidean region ``on the left'' and we have obtained
``banana-like'' structure of this region $^{8}$ (see Fig. 1c).

After the
inclusion of the small term describing the non-minimal coupling
between scalar field and gravity $(\xi \neq 0)$ we obtain the second
Euclidean region in the upper left corner of the plane $(x,a)$ $^{16}$.
This
new region is non-compact and unrestricted from above (see Fig. 1d).
While increasing the value of the parameter $\xi$ this second
Euclidean region drops down and at some value of $\xi$ .
joins with the first banana-like Euclidean region (see Fig.1e).
It is easy to
find this value of $\xi$ in the absence of self-interaction of the
scalar field:
 $$\xi = \frac{16 \pi^{2} m^{4} Q^{2}}{27 m_{P}^{4}}$$
When with the growing of the value of $\xi$ we shall have the unified
Euclidean region.  The boundary of this unified region is partially
convex, partially concave (see Fig. 1f) and after further increasing
of $\xi$ it becomes convex (see Fig. 1g).

After  inclusion of self-interaction
of the scalar field $\lambda \neq 0$ we can have, depending on the
values of the parameters  $Q, \lambda, \xi$ and $m$, various
geometrical configurations of the Euclidean regions. It is easy to
estimate the condition of closing of the Euclidean region from above:
It is
\[\xi < \left(\frac{Q \lambda}{48}\right)^{2/3}\]
If this condition is satisfied only one closed ``banana-like''
Euclidean region exists (see Fig. 1h).
In th opposite case we have two options depending on the complicated
interrelation between parameters $\xi, m$ and $\lambda$. First, one can
have
two non-connected Euclidean regions (the corresponding
picture is close to that described on Fig. 1d) :
banana-like one and ``bag-like'' Euclidean region
with an infinitely long narrow throat (the curves bounding the upper
Euclidean region are asymptotically clinging to the ordinate axis).
Second, one can have
one open above bag-like Euclidean region which again has
an infinitely long narrow throat (see Fig. 1i). Thus, we have seen
that the inclusion of the charge $Q$, non-minimal coupling $\xi \neq
0$ and self-interaction of the scalar field implies a large variety
of possible geometries of Euclidean regions in minisuperspace.

\section{Equations of motion, instantons and initial conditions for
inflation}
\hspace{\parindent}
Now using the Lagrangian (2.9) and choosing the gauge $N = 1$ we can
get the following equations of motion:
\begin{eqnarray}
&&\frac{m_{P}^{2}}{16 \pi}\left(\ddot{a} + \frac{\dot{a}^{2}}{2 a}
+ \frac{1}{2 a} - \frac{\Lambda a}{2}\right)
+\frac{\xi \dot{a}^{2} x^{2}}{4 a} + \frac{\xi \ddot{a} x^{2}}{2}
+ \xi x \dot{x} \dot{a} + \frac{\xi \dot{x}^{2} a}{2}\nonumber \\
&&+\frac{\xi x \ddot{x} a}{2} + \frac{\xi x^{2}}{4 a}
+\frac{a \dot{x}^{2}}{8}
-\frac{m^{2} x^{2} a}{8} + \frac{Q^{2}}{4 a^{5} x^{2}} -
\frac{\lambda x^{4} a}{96} = 0
\end{eqnarray}
and
\begin{eqnarray}
&&\ddot{x} + \frac{3 \dot{x} \dot{a}}{a} - \frac{6 \xi x \ddot{a}}{a}
- \frac{6 \xi \dot{a}^{2} x}{a^{2}}\nonumber \\
&&-\frac{6 \xi x}{a^{2}} + m^{2} x - \frac{2 Q^{2}}{a^{6} x^{3}}
+\frac{\lambda x^{3}}{6} = 0.
\end{eqnarray}

Besides we can write down the first integral of motion of our
dynamical system which can be obtained from the super-Hamiltonian
constraint (2.8):
\begin{eqnarray}
&&-\frac{3}{8 \pi} m_{P}^{2} a \dot{a}^{2} - 3 \xi a \dot{a}^{2}
x^{2} -6 \xi x \dot{x} \dot{a} a^{2}\nonumber \\
&&+\frac{a^{3}}{2} \dot{x}^{2} - U(a,x) = 0.
\end{eqnarray}

It is obvious that Euclidean counterparts of the
Eqs.  (4.1)--(4.3) can be obtained by the changing of sign before
terms containing time derivatives.

Numerically integrating the Euclidean analog of the system of
equations (4.1)--(4.2) we can investigate the question about the
presence of instantons. Under instantons we shall understand
solutions of Euclidean equations of motion which have vanishing
velocities $\dot{x}$ and $\dot{a}$ on the boundaries of the Euclidean
regions which are given by Eq. (3.2). In our recent letter $^{16}$ we
have investigated numerically this question and have found instanton
solutions at some configurations of the Euclidean regions. Here we
would like to complement the numerical investigation the equations
of motion of our system by some qualitative analysis.

First of all it makes sense to resolve the equations (4.1)--(4.2)
in respect with the second derivatives $\ddot{a}$ and $\ddot{x}$.
The obtained expressions look as follows:
\begin{eqnarray}
&&\ddot{a} = \frac{1}{\left(\frac{m_{P}^{2}}{16\pi}
+\frac{\xi x^{2}}{2} + 3 \xi^{2} x^{2}\right)} \times \left(-\frac
{\dot{a}^{2}}{a}\left(\frac{m_{P}^{2}}{32\pi}
+\frac{\xi x^{2}}{2} + 3 \xi^{2} x^{2}\right)\right.\nonumber \\
&&-\frac{\dot{x}^{2} a (4\xi+1)}{8} +\frac{\xi\dot{a}\dot{x}x}{2}
-\frac{m_{P}^{2}}{32 \pi a} + \frac{m_{P}^{2} \Lambda a}{32 \pi}
\nonumber \\
&&-\frac{\xi x^{2}(12\xi+1)}{4 a} + \frac{m^{2} x^{2} a
(4\xi+1)}{8} -\frac{Q^{2}(4\xi+1)}{a^{5} x^{2}}\nonumber \\
&&\left.+\frac{\lambda x^{4} a (8\xi+1)}{96}\right);
\end{eqnarray}
\begin{eqnarray}
&&\ddot{x} = \frac{1}{\left(\frac{m_{P}^{2}}{16\pi}
+\frac{\xi x^{2}}{2} + 3 \xi^{2} x^{2}\right)} \times \left(
-\frac{3 m_{P}^{2} \dot{x} \dot{a}}{16 \pi a}\right.\nonumber \\
&&-\frac{3 \xi (1+4 \xi) \dot{x} \dot{a} x^{2}}{2 a}
+\frac{3 m_{P}^{2} \xi \dot{a}^{2} x}{16 \pi a^{2}} -
\frac{3 \xi (1 + 4 \xi) \dot{x}^{2} x}{4}\nonumber \\
&&+\frac{3 m_{P}^{2} \xi x}{16 \pi a^{2}}
+\frac{3 \xi^{2} x^{3}}{2 a^{2}}
-\frac{m_{P}^{2} m^{2} x}{16 \pi}\nonumber \\
&&+\frac{\xi m^{2} x^{3}}{4} + \frac{m_{P}^{2} Q^{2}}{8 \pi a^{6}
x^{3}} - \frac{\xi Q^{2}}{2 a^{6} x}\nonumber \\
&&\left.- \frac{m_{P}^{2} \lambda x^{3}}{96 \pi}
-\frac{\lambda \xi x^{5}}{48} + \frac{3 m_{P}^{2} \xi \Lambda x}{16
\pi}\right).
\end{eqnarray}

Looking at Eq. (4.3) one can easily see that in the Euclidean region,
where $U(a,x) > 0$ it is possible to
have $\dot{a} = 0$ i.e. that the cosmological radius can achieve
extremum $a_{min}$ or $a_{max}$. It is interesting to learn which
points of Euclidean region can play role of points of minimal
contraction and which ones can be points of maximal expansion. To
understand that it is necessary to put in Eq. (4.4) $\dot{a} = 0$
and to express $\dot{x}$ as a function of $x$ and $a$ from Eq.
(4.3). We shall have
\begin{eqnarray}
&&\ddot{a} = \frac{1}{\left(\frac{m_{P}^{2}}{16\pi}
+\frac{\xi x^{2}}{2} + 3 \xi^{2} x^{2}\right)} \times \left(
-\frac{m_{P}^{2} (1 + 3 \xi)}{8 \pi a}\right.\nonumber \\
&&+ \frac{m_{P}^{2} \Lambda a (1 + 2 \xi)}{16 \pi}
-\frac{\xi x^{2} (1 + 6 \xi)}{a} + \frac{m^{2} x^{2} a (1 + 4
\xi)}{4}\nonumber \\
&&\left. -\frac{\lambda x^{4} a (1 + 6 \xi)}{48}\right).
\end{eqnarray}

Apparently, if we have at some point $\ddot{a} > 0$ this point
would correspond to the possible minimal contraction of the Universe
while $\ddot{a} < 0$ corresponds to the possible maximal expansion.
We can obtain the curve separating the points of possible minimal
contraction from those of possible maximal expansion by putting the
right-hand side of Eq.  (4.6) equal to zero.  In the simplest case
when all the parameters besides $m$ are equal to zero we reproduce
the simple hyperbolic curve
$$x = \pm\sqrt
{\frac{1}{2 \pi}} \frac{m_{P}}{m a},$$
separating the points of minimal
contraction from those of maximal expansion which was discussed in
paper $^{10}$ and in also in paper $^{11}$, where it was written down
in terms of phase space. This curve repeat that of the hyperbole
$$x = \pm\sqrt
{\frac{3}{4 \pi}} \frac{m_{P}}{m a},$$
separating the Euclidean region from the Lorentzian one and differs
from it by the multiplicative factor $\sqrt{\frac{2}{3}}$.

It is interesting to mention that the form of the curve given by Eq.
(4.6) does not depend of the charge $Q$.

One can get also an analogous curve separating the points of
possible maximum and minimum values of the absolute value of scalar
field $x$.  These points can exist only in Lorentzian region $U(x,a)
< 0$ as one sees from Eq. (4.3). Putting in eq. (4.5) $\dot{x} = 0$
and expressing $\dot{a}$ through variables $x$ and $a$ by resolving
Eq. (4.3) we have got the following equation:
\begin{eqnarray}
&&\ddot{x} = \frac{1}{\left(\frac{m_{P}^{2}}{16\pi}
+\frac{\xi x^{2}}{2} + 3 \xi^{2} x^{2}\right)
\left(\frac{m_{P}^{2}}{16\pi}
+\frac{\xi x^{2}}{2}\right)}\nonumber \\
&&\left(\frac{m_{P}^{4} \xi \lambda x}{64 \pi^{2}}
+\frac{3 m_{P}^{2} \xi^{2} x^{3}}{32 \pi a^{2}}
+\frac{3 \xi^{3} x^{5}}{4 a^{2}}\right.\nonumber \\
&&-\frac{m_{P}^{4} m^{2} x}{256 \pi^{2}}
+\frac{\xi^{2} m^{2} x^{5}}{8}
+\frac{m_{P}^{2} Q^{2}}{128 \pi^{2} a^{6} x^{3}}\nonumber \\
&&+\frac{m_{P}^{2} Q^{2} \xi}{16 \pi a^{6} x}
-\frac{\xi^{2} Q^{2} x}{4 a^{6}}
-\frac{m_{P}^{4} \lambda x^{3}}{1536 \pi^{2}}\nonumber \\
&&\left.-\frac{m_{P}^{2} \xi \lambda x^{5}}{192 \pi}
-\frac{\lambda \xi^{2} x^{7}}{96}
+\frac{3 m_{P}^{2} \xi^{2} \Lambda x^{3}}{32 \pi}\right).
\end{eqnarray}

In the simplest case when only $m \neq 0$ the separating curve has an
extremely simple form
$$ x = 0. $$
Thus all the points $x > 0$ can play the role of points of possible
maximum value of $x$.

Here it is important to add that Eqs. (4.6)--(4.7) can be used also
for the analysis of Euclidean equations of motion as well. In this
case the points of maximal expansion and minimal contraction can be
placed only in Lorentzian region while the points where $x$ can take
minimal and maximal values can exist only in Euclidean region.

Now, recalling that an instanton is the solution of the classical
Euclidean equations of motion which begins and ends at the boundary
of Euclidean and Lorentzian regions with vanishing velocities
$\dot{a}$ and $\dot{x}$ one can notice that instantonic trajectory
should cross both separating curves (in $a$ and in $x$) at least
once. Thus, studying the location of separating curves we can get
some information about possible existence and form of instantons.

One can easily see that in the case of the simplest model with
real scalar field with or without cosmological constant $\Lambda$
(see Fig. 1a) instantons cannot exist (except the trivial case
$x = 0$. However, in the case when the non-zero charge $Q$ is
switched on, both separating curves (we shall call them $x$ -
separating curve and $a$ - separating curve intersect the Euclidean
field and instanton can exist. And it really exists as the numerical
calculation confirms (see Fig. 2a). Using the end point of instanton
on the right part of the boundary of Euclidean "banana - like" region
as an initial point for the Lorentzian trajectory one can see that
this Lorentzian trajectory has quasi-inflationary behaviour
$^{8,9,17}$.

Let us now consider the model with non-minimally coupled complex
scalar field. To begin with we choose the value of $\xi$ is small
enough and we have two disconnected Euclidean regions (such situation
was presented on Fig. 1d). On the Fig. 2b depicted not only the
configuration of Euclidean regions but also the form of $a-$ and $x$-
separating curves. One can see that as in the previous case instanton
can exist in the closed "banana-like" region and it does exist and
can provide the suitable initial conditions for the beginning of
inflationary lorentzian trajectory. At the same time another solution
of Euclidean equations of motion with vanishing initial and final
velocities is present. This solution connects different Euclidean
regions and at first glance looks rather strange because it goes
through Lorentzian region while the more habitual instantons used
to intersect Euclidean regions. However, the Lorentzian trajectory
beginning from the final point of this instanton has a rather
peculiar non-inflationary behaviour and hardly can be used for the
description of quantum tunneling of the Universe from nothing.

Then while non-minimal coupling constant $\xi$ is growing the path
covered by this ``peculiar '' instanton is getting smaller and at the
moment when two region meet in one point (see Fig. 1e) this second
instanton degenerates into the point and disappear.

With further increasing of $\xi$ we have one Euclidean region open
above. In this case if $\xi$ is not very high we have two instantons
which lie inside of the Euclidean region $^{17}$ and intersect both
separating curves (see Fig. 2c). Lower instanton corresponds to the
local maximum of absolute value of action of the Euclidean
trajectories going through Euclidean region, while upper instanton
correspond to the local minimum of the absolute value of action.
It is necessary to add that taking the growing
initial values of $x$ on the boundary of the Euclidean region under
consideration (above the second instanton) we shall get the Euclidean
trajectories with the unboundedly growing action. Both the end points
of instantons can be used as initial points of Lorentzian
trajectories having quasi-inflationary behaviour (see Fig. 2c).

It is important to notice that the upper instanton corresponding to
the minimum of the absolute value of Euclidean action provides the
existence of the peak of the probability distribution in the
tunneling wave function of the Universe $^{4}$, because this function in
the tree-level approximation has the behaviour
\[ \Psi_{T} \sim \exp (-|I|).\]
At the same time the lower instanton corresponding to
maximum of the absolute value of Euclidean action provides the
existence of the peak of the probability distribution in the Hartle-
Hawking wave function of the Universe $^{3}$, having behaviour
\[ \Psi_{T} \sim \exp (+|I|).\]
Thus if we choose no-boundary proposal for the wave function of the
Universe we should use the end point of lower instanton for the
definition of the most probable initial condition for the
cosmological evolution while if we choose tunneling proposal for the
wave function of the Universe we should use the end point of upper
instanton to fix the most probable initial boundary conditions for
inflation.

As was already shown in $^{17}$ in the case when the parameter $\xi$
is large and the boundary of the Euclidean region has a convex form
(see Fig. 1g) we do not have instantons at all. It is quite
understandable now because for such a choice of parameters the
$x$-separating curve does not go through Euclidean region (see Fig.
2d) and the necessary conditions of existence of instantons are not
satisfied.  It is interesting to notice that in both these cases
(Fig. 2c and Fig. 2d) due to the indefinite increasing of the
absolute value of action with the increasing of the initial value of
scalar field $x$ at right boundary between Lorentzian and Euclidean
regions the tunneling wave function of the Universe is normalizable
already in tree-level approximation (cf. Refs. [5,7]).

Now we can investigate the instantonic solutions and initial
conditions for the inflation for the case when self-interaction of
inflaton field ($\lambda \neq 0$) is taken into account. As was
described in the preceding section three possible configurations of
Euclidean regions can exist. In the case when only one closed
Euclidean region exists (see Fig. 1h) one can find only one instanton
as usual. In the case when we have two disjoint Euclidean regions
we can discover two instantonic configurations: one
going through closed ``banana-like'' and one ``peculiar'' connecting
the boundaries of two Euclidean regions and going through Lorentzian
region. The configuration of these instantons closely resembles that
of Fig. 2b.

The most interesting is the situation when we have one
open above Euclidean region (see Fig. 1i). In this case depending on
values of parameters $\xi, \lambda, Q$ and $m$ one can observe three
different situations. In the first case depicted on Fig. 2e we have
two instantons corresponding to the minimum and maximum of absolute
value of Euclidean action. In the second case we
do not have instantons at all. For this configuration the large value
of the parameter of non-minimal coupling $\xi$ is essential.
The absolute value of action is growing monotonically with the
increasing of the initial value of scalar field $x$ on the right
boundary of Euclidean region, and tends to some fixed value (this
value is finite in contrast with the case of the scalar field without
self-interaction).

And in
the third case we have only one instantonic
solution corresponding to the maximum of absolute value of Euclidean
action. To realize this case it is necessary to have the value of
the constant of self-interaction $\lambda$ large enough to provide
the decreasing of the absolute value of action at the increasing of
initial value of $x$, however, this value of $\lambda$ must not to be
too large to escape the closing of the Euclidean region from above.

It is interesting also to consider the case when the cosmological
constant disappears ($\Lambda = 0$). The disappearance of the
cosmological constant implies the opening of the Euclidean region
on the right. The most interesting situation occurs when we have
one Euclidean region open above and from the right (see Fig. 2f)
Here instead of two instantons we have only one: that corresponding
to the minimum of the absolute value of the Euclidean action and
correspondingly to the probability peak in the tunneling wave
function of the Universe. The second ``lower'' instanton (cf. Fig.
2c) turns into the trajectory infinitely travelling through
Euclidean region without reaching its boundary with Lorentzian region
(see Fig. 2f). Thus in this case only the tunneling wave function of
the Universe can predict the most probable initial conditions for the
beginning of the cosmological evolution in contrast with the
situation when we have only one closed Euclidean region where
the Hartle- Hawking wave function of the Universe is preferable.

The received instantonic trajectories giving us the initial
conditions for the inflationary stage of cosmological evolution can
be studied from the point of view of the restrictions on the
parameters of the model which can be obtained from phenomenological
considerations. Here we shall discuss the restrictions on the
parameters of the model provided we choose the tunneling prescription
for the wave function of the Universe and we wish to have the
sufficient number of $e$-foldings (usually it is taken equal to 60).
One can show using numerical investigations that for the model without
self-interaction ($\lambda = 0$) such restrictions have the following
form:
\[\xi < 0.002;\ ;\;\; Q < \frac{50 m_{P}^{2}}{m^{2}}.\]

In the case when we have self-interaction ($\lambda \neq 0$) the
restrictions have the more complicated form. First, to provide the
beginning of the upper instanton more high it is necessary to have
the following connection between $\xi, \lambda$ and $Q$:
\[\xi \sim \frac{1}{48\pi}\frac{\lambda m_{P}^{2}}{m^2}.\]
And to escape the closing of the Euclidean  region from above one
should have
\[Q < \frac{1}{\sqrt{48\pi}} \frac{\lambda m_{P}^{2}}{m^3}.\]

\section{Generalized equations of motion and separating curves}
\hspace{\parindent}
Let us consider the cosmological model with the complex scalar field
and an arbitrary form of the interaction between this field and
gravity, arbitrary form of the potential of scalar field and
arbitrary form of the kinetic term for the scalar field as well,
The action of such a model looks like that
\begin{equation}
S = \int d^{4}x \sqrt{-g} \left(R U(\phi) + \frac{1}{2} G(\phi)
g^{\mu\nu}\phi_{,\mu}\phi_{,\nu} - V(\phi)\right).
\end{equation}
In the framework of the minisuperspace model considered in the
present paper the action (5.1) has the following form
\begin{eqnarray}
&&S = 2\pi^{2} \int dt a^{3} \left(6\left(\frac{\dot{a}^{2}}{a^{2}}
+ \frac{\ddot{a}}{a} + \frac{1}{a^{2}}\right) U(x)\right.\nonumber \\
&&\left.+\frac{1}{2}G(x)(\dot{x}^{2} + x^{2}\dot{\theta}^{2})
- V(x)\right)
\end{eqnarray}
Conservation rule for the derivative of phase $\theta$ now turns into
\begin{equation}
G(x) a^{2} x^{2} \dot{\theta} = Q.
\end{equation}
The effective Lagrangian for the
minisuperspace variables $a$ and $x$ is
\begin{eqnarray}
&&L=-6 a \dot{a}^{2} U(x) - 6 a^{2} \dot{a} \dot{x} U'(x) + 6 a U(x)
\nonumber \\
&&+\frac{G(x) a^{3} \dot{x}^{2}}{2} - \frac{Q^{2}}{2 G(x) a^{3}
x^{2}} - a^{3} V(x).
\end{eqnarray}
>From the Lagrangian (5.4) one can obtain the following equations of
motion:
\begin{eqnarray}
&&12 a U \ddot{a} + 6 U \dot{a}^{2} + 12 a U'\dot{a} \dot{x} +
6 a^{2} U''\dot{x}^{2} + 6 a^{2} U' \ddot{x} \nonumber \\
&&+ 6 U +\frac{3}{2} G a^{2} \dot{x}^{2} +\frac{3 Q^{2}}{2 G a^{4}
x^{2}} - 3a^{2} V = 0
\end{eqnarray}
and
\begin{eqnarray}
&&G a^{3} \ddot{x} + 3 G a^{2} \dot{a} \dot{x} - 6 a U' \dot{a}^{2}
-6 a^{2} U'\ddot{a}\nonumber \\
&&+\frac{G' a^{3} \dot{x}^{2}}{2} - 6 a U' -\frac{Q^{2}}{G a^{3}
x^{3}} - \frac{Q^{2} G'}{2 G^{2} a^{3} x^{2}} + a^{3} V' = 0.
\end{eqnarray}
It is more convenient to have these equation resolved in respect with
accelerations $\ddot{x}$ and $\ddot{a}$:
\begin{eqnarray}
&&\ddot{x} = \frac{1}{U G + 3 U'^{2}}\times \left(-\frac{3 G U
\dot{a} \dot{x}}{a} - \frac{6 U'^{2} \dot{a} \dot{x}} {a} + \frac{3 U
U'\dot{a}^{2}}{a^{2}}\right.\nonumber \\
&&- 3 U' U'' \dot{x}^{2} - \frac{3 G U' \dot{x}^{2}}{4} -
\frac{U G' \dot{x}^{2}}{2} + \frac{3 U U'}{a^{2}}
-\frac{3 Q^{2} U'}{4 G a^{6} x^{2}}\nonumber \\
&&\left.+\frac{3 V U'}{2} + \frac{Q^{2} U}{G a^{6} x^{3}}
+\frac{Q^{2} U G'}{2 G^{2} a^{6} x^{2}} - U V'\right);
\end{eqnarray}
\begin{eqnarray}
&&\ddot{a} = \frac{1}{U G + 3 U'^{2}}\times \left(-\frac{U G
\dot{a}^{2}}{2 a} - \frac{3 U'^{2} \dot{a}^{2}}{a} +
\frac{G U'\dot{a} \dot{x}}{2} - \frac{G U'' a \dot{x}^{2}}{2}
\right.\nonumber \\
&&-\frac{G^{2} a \dot{x}^{2}}{8} + \frac{a G' U' \dot{x}^{2}}{4}
-\frac{U G}{2 a} - \frac{3 U'^{2}}{a} -\frac{Q^{2}}{8 a^{5} x^{2}}
\nonumber \\
&&\left.+\frac{a V G}{4} - \frac{Q^{2} U'}{2 a^{5}x^{3} G}
-\frac{Q^{2} U'G'}{4 G^{2} a^{5} x^{2}} + \frac{a U' V'}{2}\right).
\end{eqnarray}
(5.8)
Now the first integral of motion of our system looks like that:
\begin{equation}
\frac{G a^{3} \dot{x}^{2}}{2} - 6 a U \dot{a}^{2} - 6 a^{2} U'
\dot{a} \dot{x} - {\cal U} = 0,
\end{equation}
(5.9)
where the superpotential ${\cal U}$ is
\begin{equation}
{\cal U} = 6 a U - a^{3} V - \frac{Q^{2}}{2 G a^{3} x^{2}}.
\end{equation}
(5.10)
The equation
\begin{equation}
{\cal U} = 0
\end{equation}
gives us the boundary between Euclidean region (${\cal U} > 0$)
and Lorentzian one (${\cal U} < 0$).

Now let us study the turning points for $a$ and $x$.
It is easy to see from Eq. (5.9) that the condition $\dot{a} = 0$
can be satisfied only if ${\cal U} > 0$ i.e. in Euclidean region.
Thus, we have physically meaningful definition of Euclidean region:
it is the part of the minisuperspace where one can find the turning
points (the points of minimal contraction $\dot{a} = 0,\ddot{a} > 0$
and the points of maximal expansion $\dot{a} = 0,\ddot{a} < 0$ for a
cosmological radius $a$).
Moreover, one can see that this property of Euclidean regions can be
generalized for the more wide class of cosmological models than
minisuperspace models consided here. The point is that the turning
points are points which correspond to extrema of the conformal factor
(i.e. cosmological radius) of
the model under consideration. However, the conformal factor i
is the only variable the squared velocity of which is included
into the Lagrangian with the negative sign. Thus, the general structure
of Wheeler-DeWitt equation as well as the general structure of the
Lagrangian imply the fact that extrema of conformal factor
are located inside the Euclidean region.

To find the curve separating the possible points of
minimal contraction from those of maximal expansion we can substitute
the condition $\dot{a} = 0$ into Eq. (5.9) and we find that
\begin{equation}
\dot{x}^{2} = \frac{2 {\cal U}}{ 2 G a^{3}}.
\end{equation}
Substituting the condition $\dot{a} = 0$ and the $\dot{x}^{2}$ from
Eq. (5.12) into Eq. (5.8) we have got
\begin{eqnarray}
&&\ddot{a} = \frac{1}{U G + 3 U'^{2}}\times \left(-\frac{6 U U''}{a}
+ V U'' a + \frac{Q^{2} U''}{2 G a^{5} x^{2}}\right.\nonumber \\
&&-\frac{2 G U}{a} +\frac{3 U U' G'}{G a} - \frac{V G' U' a}{2 G}
-\frac{Q^{2} G' U'}{2 G^{2} a^{5} x^{2}}\nonumber \\
&&\left.-\frac{3 U'^{2}}{a} -\frac{Q^{2}U'}{2 G a^{5} x^{3}}
+\frac{a U' V'}{2} + \frac{a V G}{2}\right).
\end{eqnarray}
In the region where right-hand side of Eq. (5.13) is positive and
hence $\ddot{a} > 0$  we can find the points of minimal contraction
of the Universe while the region where right-hand side of Eq. (5.13)
is negative corresponds to the points of maximal expansion.  The
investigation of the turning points for $x$ can be carried out in a
similar way. The main difference is that the points of maximum and
minimum of $x$ can be found only in Lorentzian region ${\cal U < 0}$
as might be easily deduced from Eq. (5.9). Thus the curve separating
the possible points of minimal and maximal $x$ is described by  the
equation
\begin{equation}
\ddot{x} = \frac{1}{U G + 3 U'^{2}}\times
\left(2 U' V - U'V - \frac{Q^{2} U'}{2 G a^{6} x^{2}} + \frac{Q^{2}
U}{G a^{3} x^{3}} + \frac{Q^{2} G'U}{2 G^{2} a^{6} x^{2}}\right).
\end{equation}

Now let us consider some simple particular cases.
If $G = constant, U = const$ then the equation for the curve
separating the points of minimal contraction from those of maximal
expansion has a very simple form:
\begin{equation}
a^{2} = \frac{4 U}{V}.
\end{equation}
It worth noticing that the curve separating the Euclidean region from
Lorentzian one in this case has the form
\begin{equation}
a^{2} = \frac{6 U}{V}.
\end{equation}
Thus the curve (5.15) repeats the form of the curve (5.16) with the
multiplicative factor 2/3.
The simplest case when
\[U = m_{P}^{2} / 16 \pi; V = m^{2} x^{2} /2; Q = 0\]
have already been studied in the preceding section.

The investigation of the Eq. (5.14) is of a special interest because
if we find the solutions of this equation which has the form
\begin{equation}
x = x_{0},
\end{equation}
where $x_{0}$ is a constant which is independent of $a$ then we have
separating curve which is parallel to the axis $a$ and plays role of
a real tunneling geometry $^{26}$. Indeed, we can consider the
solution of Euclidean equations of motion beginning in the point $a =
0, x = x_{0}$ with the velocities $\dot{a} = 1, \dot{x} =0$.
Arriving at the point on the boundary $(x = x_{0}, a = a_{0}(x_{0}))$
with the vanishing velocities $\dot{x} = 0, \dot{a} = 0$ this
solution can be called instanton and can be continued into the
Lorentzian region along the Lorentzian equations of motion.
To begin with let us consider the case of a real scalar field when
$$Q = 0$$.
The equation which we study has in this case a very simple form
\begin{equation}
2 U' V - U V' = 0.
\end{equation}
For the case when $U = const$ the equation (5.18) boils down to the
trivial equation
\begin{equation}
V' = 0.
\end{equation}
Now let
\[ U = \frac{m_{P}^{2}}{16 \pi} + \frac{\xi x^{2}}{2}\]
and
\[ V = \Lambda + \frac{m^{2} x^{2}}{2} + \frac{\lambda x^{4}}{24}.\]
Eq. (5.18) in this case turns into
\begin{equation}
2 \xi \Lambda x - \frac{m_{P}^{2} m^{2} x}{16 \pi}
\frac{\xi m^{2} x^{3}}{2} - \frac{m_{P}^{2}\lambda x^{3}}{96\pi} =
0.
\end{equation}
The nontrivial solution of this equation is
\begin{equation}
x = \pm \sqrt{\frac{\frac{m_{P}^{2} m^{2}}{16 \pi} - 2 \xi \Lambda}
{\frac{\xi m^{2}}{2} - \frac{m_{P}^{2}\lambda}{96\pi}}}.
\end{equation}
If we choose $\lambda = 0$ and $\Lambda = 0$ when our solution
will be
\begin{equation}
x = \pm m_{P} /\sqrt{8 \pi \xi}.
\end{equation}
It is also very interesting that we can obtain real tunneling
constant in $x$ solutions of Euclidean equations of motion which can
begin from any point on the axis $x$. Indeed, if
\begin{equation}
V = \alpha U^{2},
\end{equation}
where
$\alpha$ is an arbitrary constant then it is easy to see that the
right-hand side of the Eq. (5.18) is identically equal to zero.
The simplest realization of such a situation can be obtained by
the assumption
\[ V \sim x^{2p}, \ U \sim x^{p},\]
where $p$ is an arbitrary real number.

Moreover, in the case when $Q \neq 0$ we can get the ``horizontal''
instantonic solutions adding to the requirement (5.22)) the
additional one
\begin{equation}
G = \beta U / x^{2},
\end{equation}
where $\beta$ is an arbitrary constant.
One can check that in this case Eq. (5.18) is again satisfied
identically. However, in this case we should begin instanton solution
from the left branch of the boundary between Euclidean and Lorentzian
regions in the minisuperspace.
Apparently the choice
\[V \sim  x^{4},\ U \sim x^{2},\ G = const \]
satisfies the conditions (5.22) and (5.23)

All these horizontal instantons can be treated as examples of real
tunneling geometries and can be continued into the Lorentzian region
along the Lorentzian equations of motion. These Lorentzian
trajectories represent the so-called  ``eternal'' inflation and are
of not large interest from phenomenological point of view. However,
the little deviation from the corresponding initial conditions can
supply as with long but not eternal inflation and can be used for the
realistic scenario of the cosmological evolution.

\section{Conclusion}
\hspace{\parindent}
We have considered the cosmological model with non-minimally coupled
complex scalar inflaton field. The complexity of the scalar field is
boiled down on the level of the minisuperspace consideration to the
inclusion into the theory a new quasi-fundamental constant --
classical charge of the Universe. The presence of this constant
essentially modifies the geometry of the so-called Euclidean or
classically forbidden regions and correspondingly changes the physics
of the tunneling transition or the ``quantum birth from nothing of
the Universe''.

Moreover, consideration of such a model allows us to make some
predictions for  the most probable initial conditions for the
classical evolution of the Universe. Both the main proposal for the
wave function of the Universe -- no-boundary $^{3}$ and tunneling $^{4}$
are investigated. It is interesting that for different choices of
parameters different proposal for the wave function of the Universe
are more reasonable. Indeed, in the cases when we have only closed
euclidean region one can define the Hartle-Hawking wave function of
the Universe predicting some preferable value for initial evolution
of the Universe; in the case when we have one non-compact Euclidean
region open from above and inside this region we have two instantons
both proposals for the wave function of the Universe are plausible;
and at last if we have one non-compact Euclidean region open from
above and from the right (in the case when the cosmological constant
 $\Lambda$ vanishes) only tunneling wave function of the Universe
has the chances for the explanation of the beginning of the
inflation.\\

{\bf ACKNOWLEDGMENTS}

We are grateful to A.O. Barvinsky, R. Kolb, G.V. Lavrelashvili,
D.N. Page, V.A. Rubakov, M.V. Sazhin, A.A. Starobinsky and
A. Vilenkin for useful discussions. This work was supported by
Russian Foundation for Fundamental Researches via grants
No 96-02-16220 and No 96-02-17591
and by INTAS via project 93-3364-Ext.

\begin{description}
\item[\rm 1.]
A.D. Linde, {\it Particle Physics and Inflationary Cosmology}
(Harwood Academic, 1990) and references therein.
\item[\rm 2.]
J. Smoot et al., {\it Aph. J.} {\bf 396}, L1 (1992);
I. Strukov et al., {\it Pis'ma Astron. Zh.} {\bf 18}, 387 (1992).
\item[\rm 3.]
J.B. Hartle and S.W. Hawking, {\it Phys. Rev.} {\bf D28}, 2960
(1983); S.W. Hawking, {\it Nucl. Phys.} {\bf B239}, 257 (1984).
\item[\rm 4.]
A. Vilenkin, {\it Phys. Lett.} {\bf 117B}, 25 (1982); {\it Phys.
Rev.} {\bf D27}, 2848 (1983); {\it Phys. Rev.} {\bf D30} 509 (1984);
{\it Phys. Rev.} {\bf D37}, 888 (1988); A.D. Linde, {\it Zh. Eksp.
Teor.  Fiz.}  {\bf 87}, 369 (1984)  [{\it Sov. Phys. JETP} {\bf 60},
211 (1984)]; Ya.B.  Zeldovich and A.A. Starobinsky, {\it Pis'ma
Astron.  Zh.}  {\bf 10}, 323 (1984) [{\it Sov.  Astron. Lett.} {\bf
10}, 135 (1984)]; V.A.  Rubakov, {\it Phys. Lett.} {\bf 148B}, 280
(1984).
\item[\rm 5.] S.W.  Hawking and D.N. Page, {\it Nucl. Phys.}
{\bf B264}, 185 (1986).
\item[\rm 6.] L.P. Grishchuk and L.V.
Rozhansky, {\it Phys. Lett.} {\bf B208}, 369 (1988); {\bf B234}, 9
(1990).
\item[\rm 7.] A.O.  Barvinsky and A.Yu. Kamenshchik,
{\it Class. Quantum Grav.} {\bf 7}, L181 (1990); {\it Phys.  Lett.}
{\bf B332}, 270 (1994); A.O. Barvinsky, {\it Phys. Rep.} {\bf 230},
237 (1993); A.Yu.  Kamenshchik, {\it Phys. Lett.} {\bf B316}, 45
(1993).
\item[\rm 8.] I.M. Khalatnikov and A. Mezhlumian, {\it Phys.
Lett.} {\bf A169}, 308 (1992); I.M. Khalatnikov and P. Schiller,
{\it Phys. Lett.}  {\bf B302}, 176 (1993).
\item[\rm 9.]
L. Amendola, I.M. Khalatnikov, M. Litterio and F.  Occhionero,
{\it Phys. Rev.}  {\bf D49},
1881 (1994).
\item[\rm 10.] D.N. Page, {\it Class. Quantum Grav.} {\bf
1}, 417 (1984).
\item[\rm 11.] V.A. Belinsky, L.P. Grishchuk,
Ya.B. Zel'dovich and I.M.  Khalatnikov, {\it J. Exp. Theor. Phys.}
{\bf 89}, 346 (1985); V.A.  Belinsky and I.M. Khalatnikov, {\it J.
Exp.  Theor.  Phys.}  {\bf 93}, 784(1987).
\item[\rm 12.] J.J.
Halliwell and J.B.  Hartle, {\it Phys.  Rev.}  {\bf D41}, 1815
(1990).
\item[\rm 13.] G.V.  Lavrelashvili, V.A.  Rubakov, M.S.
Serebryakov and P.G.  Tinyakov, {\it Nucl. Phys.} {\bf B329}, 98
(1990); V.A. Rubakov and P.G.  Tinyakov, {\it Nucl. Phys.} {\bf B342},
430 (1990).
\item[\rm 14.] D.  Scialom and P. Jetzer, {\it Phys.
Rev.}  {\bf D51}, 5698 (1995).
\item[\rm 15.] B.L.Spokoiny, {\it
Phys.  Lett.}  {\bf 129B}, 39 (1984); D.S. Salopek, J.R. Bond and J.M.
Bardeen, {\it Phys.  Rev.} {\bf D40}, 1753 (1989); R. Fakir and W.G.
Unruh, {\it Phys.  Rev.}  {\bf D41}, 1783 (1990); R. Fakir, S. Habib
and W.G.  Unruh, {\it Aph.  J.}  {\bf 394}, 396 (1992); R. Fakir and
S. Habib, {\it Mod.  Phys. Lett.}  {\bf A8}, 2827 (1993).
\item[\rm 16.] A.O.  Barvinsky, A.Yu.  Kamenshchik and I.P. Karmazin, {\it
Phys. Rev.}  {\bf D48}, 3677 (1993).
\item[\rm 17.] A.Yu. Kamenshchik, I.M. Khalatnikov and A.V.
Toporensky, {\it Phys. Lett.} {\bf B357}, 36 (1995).
\item[\rm 18.] K. Lee, {\it Phys. Rev. Lett.} {\bf 61},
263 (1988).
\item[\rm 19.] S. Coleman and K. Lee, {\it Nucl. Phys.}
{\bf B329}, 387 (1990).
\item[\rm 20.] K. Lee, {\it Phys. Rev.}
{\bf D50}, 5333 (1994).
\item[\rm 21.] J.D. Brown, C.P.Burgess, A.
Kshisagar, B.F. Whiting and J.W. York, {\it Nucl.  Phys.} {\bf B326},
213 (1989); J. Twamley and D.N. Page, {\it Nucl.  Phys.} {\bf B378},
247 (1992).
\item[\rm 22.] A.O. Barvinsky and A.Yu. Kamenshchik,
{\it Phys. Rev.}  {\bf D50}, 5093 (1994);
A.O. Barvinsky, A.Yu. Kamenshchik and I.V. Mishakov,
{\it Nucl. Phys.} {\bf 491}, 387 (1997).
\item[\rm 23.] A.I. Baz',
Ya.B. Zel'dovich and A.I. Perelomov, {\it Scattering, reactions and
decays in non-relativistic quantum mechanics} (Nauka, Moscow, 1971).
\item[\rm 24.]
R. Landauer and Th. Martin, {\it Rev. Mod. Phys.} {\bf 66}, 217 (1994).
\item[\rm 25.]
A. Peres, {\it Am. J. Phys.} {\bf 48}, 552 (1980).
\item[\rm 26.]
G.W. Gibbons and J.B. Hartle, {\it Phys. Rev.}  {\bf D42}, 2458 (1990).
\end{description}
\newpage
{\bf Captions to Figures}

FIG. 1. The geometries of the Euclidean regions at different values
of the parameters: a) $Q=\Lambda=\lambda=\xi=0, m \neq 0$
b)$Q=\lambda=\xi=0, m \neq 0, \Lambda \neq 0$;
c)$\lambda=\xi=0, m \neq 0, \Lambda \neq 0, Q\neq 0 0$;
d)$\lambda=0,  m \neq 0, \Lambda \neq 0, Q\neq 0, \xi \neq  0$;
e)$\lambda=0, \xi = \frac{16 \pi^{2} m^{4} Q^{2}}{27 m_{P}^{4}} $;
f)$\lambda=0, \xi > \frac{16 \pi^{2} m^{4} Q^{2}}{27 m_{P}^{4}} $;
g)$\lambda=0, \xi \gg \frac{16 \pi^{2} m^{4} Q^{2}}{27 m_{P}^{4}} $;
h)$\lambda \neq 0, \xi < \left(\frac{Q \lambda}{48}\right)^{2/3}$;
i)$\xi > \left(\frac{Q \lambda}{48}\right)^{2/3},
 \xi > \frac{16 \pi^{2} m^{4} Q^{2}}{27 m_{P}^{4}}$.

FIG. 2. Boundaries between Euclidean and Lorentzian regions
represented by short-dashed lines, instanton trajectories
represented by bold lines, Lorentzian trajectories represented
by long-dashed lines; $x-$ and $a-$ separating curves represented
by thin lines.

\end{document}